%% file: templateArxiv.tex
\documentclass{article}
\usepackage{filecontents}
\usepackage{PRIMEarxiv}
\usepackage[utf8]{inputenc} 
\usepackage[T1]{fontenc}    
\usepackage{hyperref}       
\usepackage{url}            
\usepackage{natbib}
\usepackage{amsfonts, amsmath, amssymb, amsthm}
\usepackage{nicefrac}       
\usepackage{microtype}      
\usepackage{graphicx}
\usepackage{graphics}
\usepackage{wrapfig}
\usepackage{fancyhdr}
\usepackage{enumitem}
\usepackage{multirow}
\usepackage{mathrsfs}
\usepackage{float}
\usepackage{subfig}
\usepackage{epsfig}
\usepackage{bm}
\usepackage{footnote}
\usepackage{rotating}
\usepackage{subcaption}
\usepackage{caption}
\usepackage{xcolor}
\usepackage{geometry, tabularx, booktabs}
\usepackage{colortbl}
\usepackage{lipsum}

\title{A Class of Semiparametric Yang and Prentice Frailty Models
}

\author{
  Cássius Henrique Xavier Oliveira \\
    Department of Statistics \\
    Federal University of Minas Gerais \\
    Belo Horizonte, MG, Brazil \\
    \texttt{cassiushenrique@yahoo.coom.br} \\
  \And
  Fábio Nogueira Demarqui \\
    Department of Statistics \\
    Federal University of Minas Gerais \\
    Belo Horizonte, MG, Brazil \\
    \texttt{fndemarqui@est.ufmg.br} \\
  \AND
  Vinícius Diniz Mayrink \\
    Department of Statistics \\
    Federal University of Minas Gerais \\
    Belo Horizonte, MG, Brazil \\
    \texttt{vdm@est.ufmg.br} \\
}

\usepackage{filecontents}

\begin{document}
\maketitle

\begin{abstract}
\input chapters/abstract
\end{abstract}

\keywords{Survival \and Frailty \and Piecewise exponential distribution \and Bernstein polynomials}

\section{Introduction}
    \input chapters/1-introduction

\section{Model formulation} \label{cap:ProposedModels}
    \input chapters/3-ProposedModels

\section{Monte Carlo simulation study} \label{cap:montecarlo}
    \input chapters/4-MonteCarloStudy

\section{Real applications} \label{cap:realAplications}
    \input chapters/5-RealApplications

\section{Final remarks and future research}  \label{cap:concl}
    \input chapters/6-Conclusion

\section*{Acknowledgments}
To the Federal University of Minas Gerais and CAPES, for the financial support granted during the period of my doctorate.

\section*{Appendix A: Simulation study comparison criteria}
Consider a generic parameter, whose true value is $\Phi$, and $\hat{\Phi}_K$ the posterior estimate obtained from the $k$-th Monte Carlo replica, which $k \in \left\{1, \hdots, M_C\right\}$. The average estimate (est) is given by $$\text{est}(\Phi) = \frac{1}{M_C} \sum_{k = 1}^{M_C} \hat{\Phi}_k.$$ It is possible to compute bias by $$\text{RB} (\%) = \frac{100}{M_C} \sum_{k=1}^{M_C} \frac{\hat{\Phi}_k - \Phi}{|\Phi|}.$$ Additionally, we can compute the average standard error (ASE) of the estimates by $$\text{ASE} = \frac{1}{M_C} \sum_{k=1}^{M_C} \text{se}(\hat{\Phi}_k),$$ where $\text{se}(\hat{\Phi}_k)$ represents the mean of standard error estimates of $\Phi$. We are also interested in evaluating the standard deviation estimate (SDE) de $\Phi$ by $$\text{SDE} = \left\{\frac{1}{M_C-1} \sum_{k=1}^{M_C} \left[\hat{\Phi}_k - \text{est}(\Phi) \right]^2 \right\}^{1/2}.$$ In a well-fitted model, we note these characteristics: est$(\Phi)$ should be close to the true value $\Phi$; SDE and ASE should be similar; RB(\%) should approximate zero; and CP should be close to the pre-defined confidence level ($1 - \alpha$). When the ASE $<$ SDE, is expected CP $ < 1 - \alpha$. On the other hand, if ASE $>$ SDE, it is expected that CP $ > 1 - \alpha$.

\section*{Appendix B: Widely Applicable Information Criterion}
The Widely Applicable Information Criterion (WAIC) calculates the log-likelihood of the data given a model, denoted as $\widehat{\mbox{lppd}}$, and penalizes for the complexity of this model, considering the effective number of parameters. Lower WAIC values indicate a model with a better predictive fit. Unlike AIC, WAIC is based on a weighted average of all parameter posterior distributions rather than just a point estimate, which can provide a more robust assessment of model quality \citep{loo}. Mathematically, it is an alternative approach to estimating the expected log pointwise predictive density and is calculated by
\begin{align}
    \widehat{\mbox{elppd}}_{\text{waic}} = \widehat{\mbox{lppd}} - \widehat{p}_{\text{waic}}, \nonumber
\end{align}
where $\widehat{p}_{\text{waic}}$ is the estimated effective number of parameters and is computed based on the sum of the posterior variance of the log-likelihood function. In practical terms, we can calculate using the posterior variance of the log predictive density for each data point $y_i$, i.e., $$\widehat{p}_{\text{waic}} = V_{s=1}^S \left[\log p \left(y_i | \Theta^{(s)} \right) \right],$$ in which $$V_{s=1}^S a_s = \frac{1}{S} \sum_{s=1}^S (a_s - \Bar{a})^2.$$
Then, we define 
\begin{align}
    \text{WAIC} = -2 ~ \widehat{\mbox{elppd}}_{\text{waic}}. \nonumber
\end{align}

To compare the quality of the fits of our models, we applied the function \texttt{waic} available in the package \texttt{loo} \citep{vehtari2021package}.

\bibliographystyle{apalike}
\bibliography{references}

\end{document}

%% file: chapters/abstract.tex
The Yang and Prentice (YP) regression models have garnered interest from the scientific community due to their ability to analyze data whose survival curves exhibit intersection. These models include proportional hazards (PH) and proportional odds (PO) models as specific cases. However, they encounter limitations when dealing with multivariate survival data due to potential dependencies between the times-to-event. A solution is introducing a frailty term into the hazard functions, making it possible for the times-to-event to be considered independent, given the frailty term. In this study, we propose a new class of YP models that incorporate frailty. We use the exponential distribution, the piecewise exponential distribution (PE), and Bernstein polynomials (BP) as baseline functions. Our approach adopts a Bayesian methodology. The proposed models are evaluated through a simulation study, which shows that the YP frailty models with BP and PE baselines perform similarly to the generator parametric model of the data. We apply the models in two real data sets.

%% file: chapters/1-Introduction.tex
In Survival analysis, a common goal is to evaluate potential risk factors on the occurrence of events. Proportional hazards (PH) regression models, such as \citet{cox1972}, and proportional odds (PO) model \citet{bennett1983} are possible approaches. Another alternative is the Yang and Prentice (YP) regression model \citet{yang2005}, which includes the PH and PO models as particular cases. In YP models, the survival functions are allowed to intersect and this provides an advantage over the PH and PO models \citet{demarqui2021}. Several works address the YP model in the literature. \citet{diao2013efficient} studied the extension of the YP model to accommodate potentially time-dependent covariates. \citet{wang2013} developed methods to calculate the sample size based on YP models. \citet{demarqui2019} introduced an approach to fit the YP model using Bernstein polynomials (BP) for handling the baseline hazard and odds, applicable within both frequentist and Bayesian frameworks. \citet{demarqui2021} proposed a semiparametric model for survival data using the YP regression model and the piecewise exponential  (PE) distribution as the baseline hazard function. The fit of this model can be done using \texttt{R} package \texttt{YPPE} \citep{yppeManual}. \citet{walmirtese} proposed a class of multivariate survival models based on Archimedean copulas with margins modeled by the YP model.  

Due to the flexibility of BP in approximating continuous functions, some works in the literature bring applications of these polynomials in survival analysis. \citet{panaro2020} developed an \texttt{R} package named \texttt{spsurv} \citet{spsurvManual} model survival times using BP coupled to some regression structures as PH and PO models. \citet{demarqui2019} used the BP to model the baseline functions of the YP model.

When it comes to the construction of survival models, in various research, it is commonly assumed that the times until the event are mutually independent. However, this assumption may not hold in certain scenarios. Consider, for instance, a patient who is infected repeatedly by a virus. We say that the individual experienced recurrent events of infections. Assuming independence in these cases may be inappropriate \citep{klein2006}. Immunity acquired due to an infection can change the likelihood of future infections. Therefore, the recurrent events are not mutually independent, as one event can alter susceptibility to subsequent events.

When survival times exhibit correlations among them, the data are classified as multivariate. This scenario arises, for example, in cases of recurrent events experienced by individuals. Conversely, when the assumption of independence between time-to-events holds true, the data are considered univariate. \citet{klein2006}, \citet{hanagal2011} and others argue that a commonly used approach to deal with some dependence on survival data is to assume that the time-to-event is conditionally independent on a set of unobserved variables, called frailties. The concept of frailty, introduced by \citet{vaupel1979}, defines it as a latent and multiplicative random variable. The authors used frailties, also called random effects, to explain the effect of unobserved heterogeneity on the mortality of a population. \citet{clayton1985} used frailties to explain the heterogeneity about the hazard function in an extension of the PH model for multivariate survival data. Frailty models can also be used to accommodate the association between recurrent events, as in \citep{lawless1987}.
\citet{huang2004} considered a joint modeling of recurrent events and a terminal event, utilizing frailties to model the correlation between the intensity of the recurrent event process and the hazard of the failure time. \citet{liu2004shared} considered frailty PH models for the recurrent and terminal event processes in which shared frailty is included in both hazard functions. \citet{mazroui2012general} proposed a joint frailty model to analyze recurrences and death, using two gamma-distributed frailties to handle both the recurrences dependence and the dependence between the recurrences and the death times. \citet{schneider2020} used the frailty model to fit survival data subjected to dependent censoring. Some works use frailty in PO models such as \citet{economou2007parametric}, \citet{lin2011bayesian}, and \citet{gupta2014}. 

The primary objective of this study is to develop a class of YP frailty models, within the Bayesian framework, with three baseline functions - exponential, PE, and BP. It allows one to analyze survival data arranged in two configurations. 
\begin{itemize}
    \item Univariate survival datasets. The frailty term serves to explain unobserved heterogeneities, that is, variations in survival time that are not explained by the fixed effects of the models. In this case, we refer to the element of the frailty as individual frailty; 
    \item Multivariate survival datasets in which individuals present recurrent events. Thus, frailty is used to accommodate the association between the survival times of the same individual. In this context, we can understand the individual as a cluster, and frailty here is referred to as shared frailty.
\end{itemize}

This work is organized as follows. Section \ref{cap:ProposedModels} presents the fundamental concepts of the YP model. Furthermore, it discusses the BP and the PE model, which will be used to model the baseline hazard functions, and presents the proposed models. Section \ref{cap:montecarlo} analyses the results of the Monte Carlo study. Section \ref{cap:realAplications} illustrates two real applications of our models. We close this text with discussions of some results and perspectives for future research in Section \ref{cap:concl}.

%% file: chapters/3-ProposedModels.tex

This section presents the YP frailty models whose baseline functions are modeled using exponential distribution, BP, and PE distribution. Let's start by discussing some theoretical aspects of the YP model. 
 \citet{yang2005} proposed a model in which survival curves can intersect. Let $T > 0$ be the random variable denoting the time-to-event. The YP model can be characterized in terms of the survival function
\begin{align}
    S(t|\mathbf{x}) = \left[1 + \frac{\nu}{\xi} R_0(t)\right]^{-\xi} \nonumber,
\end{align}
where $\nu = \exp(\mathbf{x} \pmb{\psi})$ and $\xi = \exp(\mathbf{x} \pmb{\phi})$, $\pmb{\psi} = (\psi_1,\hdots,\psi_p)'$ and $\pmb{\phi} = (\phi_1,\hdots,\phi_p)'$ are vectors of regression parameters without intercepts and $\mathbf{x} = (x_1, \hdots, x_p)$ is a row vector of explanatory variables. The function $R_0(t)$ is the baseline odds function and is defined as $R_0(t) = \frac{F_0(t)}{S_0(t)}$,  where $S_0(t) = 1 - F_0(t)$ is the baseline survival function, and $F_0(t)$ is baseline cumulative distribution function. We can express the hazard function of this model as
\begin{align}
    h(t|\mathbf{x}) = \frac{\nu \xi}{\nu F_0(t) + \xi S_0(t)} h_0(t), \nonumber
\end{align}
where $h_0(t) = -\frac{d}{dt} \log[S_0(t)]$ is baseline hazard function. The parameter $\nu$ is interpreted as the short-term hazards ratio, because
$$\lim_{t \to 0} \frac{h(t|\mathbf{x})}{h(t|\mathbf{0})} = \nu,$$
and $\pmb{\psi}$ as the short-term regression coefficients vector. 
We can interpret $\xi$ as the long-term hazards ratio, since $$\lim_{t \to \infty} \frac{h(t|\mathbf{x})}{h(t|\mathbf{0})} = \xi,$$
and $\pmb{\phi}$ is the long-term regression coefficients vector.  

The YP model can be reduced to the PH and PO models. Note that, when $\pmb{\psi} = \pmb{\phi}$, 
$h(t|\mathbf{x}) = h_0 (t) \exp(\mathbf{x} \pmb{\phi})$,
and this is the hazard function of the PH model. On the other hand, if $\pmb{\phi} = \textbf{0}$, we have
$$S (t|\mathbf{x}) = \left[1 + R_0 (t) \exp(\mathbf{x} \pmb{\psi})\right]^{-1} \Rightarrow R(t|\mathbf{x}) = \frac{F(t|\mathbf{x})}{S(t|\mathbf{x})},$$
and this is the expression of the odds function in the PO model. In YP model, when $\psi_j \phi_j < 0$, for any pair of coefficients $(\psi_j,\phi_j)$, with $j \in \{1,\hdots,p\}$, the survival curves intersect.

We establish the YP frailty model by multiplying expressions of $\nu$ and $\xi$ by a random effect $z = \exp(w)$. \citet{klein2006} highlight that $w$ is usually assumed to have a distribution with mean zero and unknown variance $\sigma^2_w$. Consider individuals $k$ and $k'$ of which  $\mathbf{x}_{k} = (x_{k,1},\hdots, x_{k,p})$ and $\mathbf{x}_k' = (x_{k',1},\hdots, x_{k',p})$ are row vectors of covariates, respectively. The ratio between the hazard functions depends not only on the observed characteristics but also on the random effects of the two individuals, since
\begin{align}
        \lim_{t \to 0} \frac{h(t|\mathbf{x}_{k}, z_k)} {h(t|\mathbf{x}_{k'}, z_{k'})} = 
        &\frac{z_k}{z_{k'}} \exp \left[\psi \left(\mathbf{x}_{k} - \mathbf{x}_{k'} \right) \right], \text{ and } \nonumber\\
        \lim_{t \to \infty} \frac{h(t|\mathbf{x}_{k}, z_k)} {h(t|\mathbf{x}_{k'}, z_{k'})} = 
        &\frac{z_k}{z_{k'}} \exp \left[\phi \left(\mathbf{x}_{k} - \mathbf{x}_{k'} \right) \right]. \nonumber
\end{align}

We model the baseline function using BP. The BP is a linear combination of bases introduced by \citet{bernstein1912}. Consider the continuous function $C(\cdot)$ defined in a range $(0, \tau]$. It can be approximated arbitrarily by BP \citet{lorentz1986}. The BP of degree $m$ evaluated in $t \in (0,\tau]$ with base $B_m = (B_{1,m},B_{2,m},\hdots, B_{m,m})$ and coefficients $b_m = (b_{1,m}, b_{2,m}, \hdots, b_{m,m})$ to approximate the function $C(\cdot)$ is $$B^C_m(t) = \sum_{k=1}^m b_{k,m} B_{k,m}(t),$$ where $b_{k,m} = C \left(\frac{k \tau}{m} \right)$ and $B_{k,m} = {m \choose k} \left(\frac{t}{\tau} \right)^k \left(1-\frac{t}{\tau} \right)^{m-k}$; $k = 1, 2, \hdots, m$. The derivative of BP with respect to $t$ can be written as $$b_m^{C}(t) = \sum_{k=1}^m \left\{C \left(\frac{k}{m}\tau \right) - C \left(\frac{k-1}{m}\tau \right) \right\} \frac{f_{\beta}\left(\frac{t}{\tau}, m-k+1 \right)}{\tau},$$ where $f_{\beta} \left (\frac{t}{\tau}, k, m-k + 1 \right)$ is the density of a Beta distribution with parameters $k$ and $m-k + 1$ valued at $\frac{t}{\tau}$. \citet{lorentz2012bernstein} shows that $B^C_m(t) \to_{m \to  \infty} C(t)$ uniformly, and $b^C_m(t) \to_{m \to  \infty} \frac{d}{dt}C(t)$ uniformly on $(0; \tau]$. 

In Survival analysis, \citet{osman2012} used the expression of $b^C_m(t)$ to model the hazard function $h(t|\pmb{\gamma})$ and $B^C_m(t)$ to handle the accumulated hazard function $H(t|\pmb{\gamma})$. Following the reasoning of these authors, assume $\pmb{\gamma} = (\gamma_1,...,\gamma_m)$, with $\gamma_k = C^* \left(\frac{k}{m}\tau \right) - C^*\left(\frac{k - 1}{m}\tau \right); \gamma_k \geq 0, k = 1,...,m$. Note that $\pmb{\gamma}$ is not time-dependent and its values are unknown. Also consider $\mathbf{g}_m(t) = \left(g_{1,m}(t),...,g_{m,m}(t)\right)'$, where $g_{k,m} (t) = \frac{1}{\tau} f_{\beta} \left(\frac{t}{\tau},k,m-k+1\right); g_{k,m} (t) \geq 0, k = 1,...,m$. 
Thus, $b_m^{C^*}(t) = \pmb{\gamma} ~ \textbf{g}_{m}(t)$.
\citet{osman2012} used this expression to model the hazard function. That is, $h(t|\pmb{\gamma}) = \pmb{\gamma} ~ \mathbf{g}_m(t); t \in [0,\infty)$.

BP has some advantages because offers flexibility in modeling different shapes of hazard functions, facilitating the model's adaptation to the specific characteristics of the data. Furthermore, BP has good derivation properties, and the log-likelihood function has a friendly form. The monotonicity of the accumulated hazard function is naturally modeled by BP, since that $\gamma_k \geq 0, \forall ~ k \in \{1,...,m\}$. This function is expressed by $H(t|\pmb{\gamma}) = \int_0^t  h(u|\pmb{\gamma}) du = \pmb{\gamma} ~ \mathbf{G}_m(t),$ with $\mathbf{G}_m(t) = \left(G_{1,m}(t),...,G_{m,m}(t)\right)',$ where $G_{m,k}(t) = \int_0^t f_{\beta}\left(\frac{u}{\tau}; k, m-k+1\right) d\left(\frac{u}{\tau}\right), \forall ~ k \in \{1,...,m\}.$ The function $G_{m,k}(t)$ is the Beta cumulative distribution function with parameters $k$ and $m-k + 1$. 

These authors also discuss some aspects of choosing $\tau$. It is necessary that $\tau < \infty$, such that $\tau = \inf\{t: S(t) = 0\}$. In practice, in survival analysis, $\tau$ is chosen as the maximum value among the times observed until the occurrence of the event of interest or until the follow-up stops. Here, we will denote it by $\hat{\tau}$. But, using this choice, it is not possible to satisfy $H(\tau| \pmb{\gamma}) = \infty$. Besides, there is no information about survival times in the region $t > \hat{\tau}$ \citet{demarqui2019}. Therefore, this choice requires an adjustment in the hazard and cumulative hazard functions. As a solution, \citet{osman2012} suggest some alterations in these functions, as follows:
\begin{align}
    h^*(t|\pmb{\gamma}) &= 
    \begin{cases}
        h(t|\pmb{\gamma}), \mbox{ if } 0 \leq t < \hat{\tau}, \\
        m ~ \frac{{\gamma}_m}{\hat{\tau}},  \mbox{ if } t \geq \hat{\tau},
    \end{cases} \nonumber \\
    H^*(t|\pmb{\gamma}) &= 
    \begin{cases}
        H(t|\pmb{\gamma}), \mbox{ if } 0 \leq t < \hat{\tau}, \\
        H(t|\pmb{\gamma}) + m(t - \hat{\tau}) ~ \frac{{\gamma}_m}{\hat{\tau}}, \mbox{ if } t \geq \hat{\tau}.
    \end{cases} \nonumber
\end{align}


Another way also found in the literature to model baseline function is the use of the PE distribution. This model was introduced by \citet{kalbfleisch1973}. Consider a time grid $\rho = \{\rho_0,\hdots,\rho_m\}$. Note that $\rho$ makes a partition of the time axis in $m$ intervals at the points $\rho_0, \rho_1, \hdots, \rho_m$, with $0 = \rho_0 <\rho_1 <\hdots <\rho_m < \infty.$ The intervals generated from that partition are $I_1 = (\rho_0, \rho_1], I_2 = (\rho_1, \rho_2], \hdots, I_m = (\rho_{m-1}, \rho_m]$. The set $\rho$ (or $m$, alternatively) can be established in different ways. \citet{breslow1974} and \citet{demarqui2021} assume that $\rho$ is a known set composed by each of different time-to-event observations. \citet{kalbfleisch1973} declare that the choice of $\rho$ can be independent of the data set. On the other hand, \citet{demarqui2021} argues that large $m$ values can provide unstable estimates. In other approaches, $\rho$ is treated as being random; see \citet{demarqui2011modeling, demarqui2012}.

In truth, the choice of $\rho$ influences the inferential results since we assume that the hazard function in each interval is constant and given by
$h(t|\gamma_j) = \gamma_j, \mbox{ for } t \in I_j, j = 1,\hdots,m \mbox{ and } \gamma_j > 0.$ The cumulative hazard function is given by $H(t|\pmb{\gamma}) = \sum_{j=1}^m \gamma_j(t_j - \rho_{j-1}),$ where
\begin{align}
    t_j = 
    \begin{cases}
        \rho_{j-1} \mbox{, if } t < \rho_{j-1}; \\
        t \mbox{, if } \rho_{j-1} < t \leq \rho_j; \\
        \rho_j \mbox{, if } t > \rho_j,
    \end{cases} \nonumber
\end{align}
for $j = 1,\hdots,m.$ From the calculation of the cumulative hazard function, we can use the expression $S(t|\pmb{\gamma}) = \exp\{-H(t|\pmb{\gamma})\}$ to find the survival function.

One of the main advantages of PE is its flexibility in adjusting to different shapes of the hazard function over time. By dividing the study period into intervals and assuming a constant hazard function within each interval, but allowing these rates to vary between the intervals, the model can adapt to a variety of hazard functions that would not be well captured by an exponential model. Furthermore, the estimated parameters for each time interval can be directly interpreted as hazard functions, making the model intuitive for researchers.

Once we have discussed the necessary elements to construct the class of models that we propose in this work, we now proceed with the establishment of the likelihood function. Let $L$ be the number of individuals. Denote by $C_i$ the time to the administrative censoring, that is, the time until loss of follow-up for some reason external to the study,  with $i = 1,\hdots, L$. Denote by $R_{i,j}$ the gap-time between the $(j-1)$-th and $j$-th occurrences of the recurrent event, and let $T_{i,j} = \sum_{j'=1}^{j} R_{i,j'}$ be the total observation time until the $j$-th recurrent event. Suppose the $i$-th subject experiences a total of $n_i$ recurrent events. When $j = n_i + 1$, $R_{i, n_i + 1} = C_i - \sum_{j=1}^{n_i} R_{i,j}$, which can be interpreted as the gap-time between the $n_i$-th recurrent event and the end of follow-up. Define $\delta_{ij} = I(T_{i,j} < C_i)$ is the failure state indicator for the $j$-th recurrent event. When $\delta_{i,j} = 0$, it indicates that the observed time is an administrative censoring time.

Assuming that the survival times $R_{i,j}, \hdots, R_{L,n_L}$ are mutually independent conditioned on the frailty term $w_i$, we can obtain the likelihood function as:
 \begin{align*}
        \mathcal{L}(\Theta| \mathcal{D}, \mathbf{w}) 
        &= \prod_{i=1}^L \left \{S(r_{i,n_i+1}|w_i)  \prod_{j=1}^{n_i} f(r_{i,j}|w_i) \right \},   \nonumber
    \end{align*}
where $\mathcal{D}$ as the set of observed data, such that $\mathcal{D} = \left\{r_{i,j}, r_{i,n_i+1}, \delta_{i,j}, \mathbf{x}_{i,j}; i = 1,\hdots, L; j = 1,\hdots, n_i\right\}.$ Let 
$\Theta = \left\{\pmb{\gamma}, \pmb{\psi}, \pmb{\phi}, \sigma^2_w\right\}$
denote the set of parameters to be estimated in the models.

%% file: chapters/4-MonteCarloStudy.tex
The simulation study was conducted to evaluate two scenarios: ($\mathcal{S}_1$) individual frailty $\left(n_i = 1\right)$, and ($\mathcal{S}_2$) shared frailty, where individuals experience recurrent events $\left(n_i \geq 1\right)$; $\forall i \in \{1,\hdots, L\}$. The data generation and model fitting were executed in the \texttt{R} programming language \citep{manualR}. We utilized \texttt{rstan} package \citet{rstanMain} to generate four Markov Chain Monte Carlo (MCMC) chains for each parameter, each chain comprising 2000 iterations, with 1000 warm-up iterations. This approach yielded posterior sample sizes of 4000 for each parameter.

We generated $M_C = 250$ Monte Carlo replicas, each one with $L = 300$ individuals using a YP model and an exponential baseline distribution (YP$_{EX}$) with rate parameter $\gamma = 1.2$. For the individual $i$, we generated the frailty $w_i \sim \mbox{Normal}~(0,1)$ and two covariates $X_{i,1} \sim \mbox{Bernoulli}~(0.5)$ and $X_{i,2} \sim \mbox{Normal}~(0,1)$. We generate the administrative censoring $C_i \sim U(0,5)$. The gap time of the $j$-th recurrence was generated by applying the inverse of the survival function, represented by $r_{i,j} = S^{-1}(u_{i,j}|\mathbf{x}_i, w_i)$, where $U_{i,j} \sim U(0,1)$. See \citet{cassius2024} for more details on how to get $S^{-1}(u|\mathbf{x}, w_i)$. In $\mathcal{S}_1$, we generate $r_{i,j}$ only once for each $i$. In contrast, in $\mathcal{S}_2$, the process of obtaining $r_{i,j}$ is repeated as long as $t_{i,j} \leq c_i$. In both scenarios, the true values of the parameters established are $\phi_1 = -1$, $\phi_2 = 2$, $\psi_1 = 2$, $\psi_2 = 2$.

The following prior distributions were used: $\pmb{\psi}, \pmb{\phi} \sim \mbox{Normal} \left(0,4^2 \right)$, $\pmb{\gamma} \sim \mbox{LogNormal}~(0,2)$, and $\sigma_w \sim \mbox{Gamma}~(0.1, 0.1).$ The prior distributions established for the regression coefficients $\pmb{\psi}$, $\pmb{\phi}$, and the the standard deviation of the frailty $\sigma_w$ are weakly informative \citep{stan}. For the parameters $\pmb{\gamma}$, we choose a prior distribution that provides greater stability in the inferential process, as suggested by \citep{demarqui2019}. Regarding the estimation of the parameters for the YP$_{BP}$ model, the choice of the BP degree is motivated by \citep{osman2012}.  These authors suggest a range of values, within which we consider $m = L^{0.4}$ in Scenario $\mathcal{S}_1$, and $m = \left[\sum_{i=1}^L n_i\right]^{0.4}$ in Scenario $\mathcal{S}_2$. We use the same criteria to define the value of $m$ in models with baseline PE distribution. All simulation study results are also available in \href{https://cassiushenrique.shinyapps.io/appSimulationsFrailty/}{cassiushenrique.shinyapps.io/appSimulationsFrailty}. The statistics used to evaluate the simulation study are described in Appendix A.

We start by evaluating the Scenario $\mathcal{S}_1$. In simulated data, approximately $68.4\%$ of the individuals experienced the event of interest, on average. The Monte Carlo summary statistics in Table \ref{tab:summaryYPFra} compare the performance of the models. The estimates from our models are close to the true values. The estimated values of $\sigma_w$ and $\phi_2$ by the YP$_{BP}$ model are slightly less accurate, deviating more from their true values, compared to the corresponding estimates obtained from the other models. The generator model (YP$_{EX}$) presented smaller biases for the short-term effect coefficients ($\psi_1$ and $\psi_2$). The YP$_{EX}$ and YP$_{PE}$ models show greater RB for the parameters of dichotomous variables in comparison to continuous covariates effects. For all the models, credibility intervals are similar. Furthermore, the ASE values are close to SDE, and CP values are close to the desired level of $0.95$, indicating good performance. 

\begin{table}
\centering
\caption{Scenario $\mathcal{S}_1$ - Monte Carlo summary statistics of the YP$_{EX}$, YP$_{PE}$, and YP$_{BP}$ with individual frailties, for $L = 300$ and $M_C = 250$.}
\label{tab:summaryYPFra}
\small
\begin{tabular}[ht]{ccrrrrrrrr}
\toprule
\multicolumn{7}{l}{ } & \multicolumn{2}{c}{95\% CI} & \\
\cmidrule(l){8-9} 
fitted model & par & true & est & RB (\%) & ASE & SDE & LW & UP & CP \\ 
\midrule
  YP$_{EX}$  & $\phi_1$ &  -1 & -0.9406 & 5.9395 & 0.2970 & 0.2935 & -1.4732 & -0.3098 & 0.9360 \\ 
   & $\phi_2$ &   2 & 2.0795 & 3.9738 & 0.3404 & 0.3388 & 1.4428 & 2.7755 & 0.9440 \\ 
   & $\psi_1$ &   2 & 2.0175 & 0.8726 & 0.2941 & 0.2940 & 1.4410 & 2.5919 & 0.9560 \\ 
   & $\psi_2$ &   2 & 2.0094 & 0.4704 & 0.1713 & 0.1713 & 1.6731 & 2.3453 & 0.9640 \\ 
   & $\sigma_w$ &   1 & 0.9971 & -0.2909 & 0.1763 & 0.1763 & 0.6489 & 1.3418 & 0.9400 \\ 
   \midrule
  YP$_{PE}$ & $\phi_1$ &  -1 & -0.9399 & 6.0081 & 0.3085 & 0.3049 & -1.4925 & -0.2813 & 0.9520 \\ 
   & $\phi_2$ &   2 & 2.0681 & 3.4029 & 0.4606 & 0.4595 & 1.2328 & 3.0187 & 0.9360 \\ 
   & $\psi_1$ &   2 & 2.1097 & 5.4830 & 0.3482 & 0.3452 & 1.4506 & 2.8150 & 0.9600 \\ 
   & $\psi_2$ &   2 & 2.0802 & 4.0083 & 0.2497 & 0.2481 & 1.6254 & 2.6026 & 0.9360 \\ 
   & $\sigma_w$ &   1 & 0.9944 & -0.5598 & 0.3186 & 0.3185 & 0.4260 & 1.6263 & 0.9320 \\ 
   \midrule
  YP$_{BP}$  & $\phi_1$ &  -1 & -0.9632 & 3.6782 & 0.3076 & 0.3063 & -1.5138 & -0.3090 & 0.9520 \\ 
   & $\phi_2$ &   2 & 2.1899 & 9.4949 & 0.5176 & 0.5086 & 1.2725 & 3.2830 & 0.9240 \\ 
   & $\psi_1$ &   2 & 2.0925 & 4.6234 & 0.3439 & 0.3418 & 1.4595 & 2.8115 & 0.9640 \\ 
   & $\psi_2$ &   2 & 2.0851 & 4.2546 & 0.2574 & 0.2556 & 1.6606 & 2.6741 & 0.9640 \\ 
   & $\sigma_w$ &   1 & 1.0628 & 6.2832 & 0.3346 & 0.3306 & 0.4848 & 1.7816 & 0.9440 \\ 
\bottomrule
\end{tabular}
\end{table}


Now we consider the Scenario $\mathcal{S}_2$ whose results are Table \ref{tab:summaryYPFraShared}. The YP$_{EX}$ shows a high degree of accuracy in estimating the parameters. The estimates for all the parameters are very close to their true values, with relative biases (RB). In contrast, when the YP$_{PE}$ is fitted, the relative biases for the same parameters increase slightly, in magnitude, except for $\sigma_w$. The YP$_{BP}$ demonstrates a further increase in relative biases for parameters such as $\phi_2$. Additionally, the ASE and SDE estimates are close to each other. The CP values are approximately $0.95$, deviating by no more than 0.026 from this level in all cases. Notably, across all models, the estimation of the frailty parameter $\sigma_w$ is consistent and accurate, with relative biases under 2\%. This indicates that all three models are reliable in capturing the shared frailty component, which is an essential aspect.

\begin{table}
\centering
\caption{Scenario $\mathcal{S}_2$ - Monte Carlo summary statistics of the YP$_{EX}$, YP$_{PE}$, and YP$_{BP}$ with shared frailties, for $L = 300$ and $M_C = 250$.}
\label{tab:summaryYPFraShared}
\small
\begin{tabular}[ht]{ccrrrrrrrr}
\toprule
\multicolumn{7}{l}{ } & \multicolumn{2}{c}{95\% CI} & \\
\cmidrule(l){8-9} 
fitted model & par & true & est & RB (\%) & ASE & SDE & LW & UP & CP \\ 
\midrule
  YP$_{EX}$ & $\phi_1$ &  -1 & -0.9831 & 1.6948 & 0.2586 & 0.2583 & -1.4664 & -0.4492 & 0.9760 \\ 
   & $\phi_2$ &   2 & 2.0664 & 3.3213 & 0.2951 & 0.2940 & 1.5314 & 2.6889 & 0.9560 \\ 
   & $\psi_1$ &   2 & 1.9834 & -0.8318 & 0.2090 & 0.2090 & 1.5739 & 2.3930 & 0.9640 \\ 
   & $\psi_2$ &   2 & 2.0076 & 0.3822 & 0.1377 & 0.1377 & 1.7418 & 2.2819 & 0.9520 \\ 
  & $\sigma_w$ &   1 & 1.0172 & 1.7159 & 0.0911 & 0.0908 & 0.8519 & 1.2085 & 0.9600 \\ 
\midrule
  YP$_{PE}$ & $\phi_1$ &  -1 & -0.9692 & 3.0824 & 0.2676 & 0.2666 & -1.4629 & -0.4144 & 0.9760 \\ 
   & $\phi_2$ &   2 & 2.0885 & 4.4248 & 0.3237 & 0.3218 & 1.5101 & 2.7747 & 0.9640 \\ 
   & $\psi_1$ &   2 & 1.9628 & -1.8625 & 0.2128 & 0.2125 & 1.5461 & 2.3796 & 0.9600 \\ 
   & $\psi_2$ &   2 & 1.9958 & -0.2118 & 0.1387 & 0.1386 & 1.7282 & 2.2718 & 0.9440 \\ 
   & $\sigma_w$ &   1 & 1.0133 & 1.3316 & 0.0906 & 0.0904 & 0.8491 & 1.2037 & 0.9600 \\ 
\midrule
  YP$_{BP}$ & $\phi_1$ &  -1 & -0.9755 & 2.4497 & 0.3115 & 0.3109 & -1.5326 & -0.3207 & 0.9560 \\ 
   & $\phi_2$ &   2 & 2.1264 & 6.3217 & 0.3226 & 0.3186 & 1.5465 & 2.8140 & 0.9640 \\ 
   & $\psi_1$ &   2 & 1.9396 & -3.0183 & 0.2435 & 0.2426 & 1.4703 & 2.4252 & 0.9360 \\ 
   & $\psi_2$ &   2 & 1.9755 & -1.2260 & 0.1447 & 0.1446 & 1.6976 & 2.2649 & 0.9440 \\ 
   & $\sigma_w$ &   1 & 1.0099 & 0.9855 & 0.0938 & 0.0937 & 0.8401 & 1.2073 & 0.9720 \\ 
\bottomrule
\end{tabular}
\end{table}



%% file: chapters/5-RealApplications.tex
To illustrate the application of our proposal, we use two databases \texttt{readmission} and \texttt{diarrhea}. The dataset \texttt{readmission} from the \texttt{frailtypack} package \citet{rondeau2012frailtypack}, was previously applied in the study of \citet{gonzalez2005sex}. These data consist of times between hospital readmissions and time to death for patients with colorectal cancer. Here, we are interested in modeling only the times until terminal events (death). In this application, we use a frailty to handle unobserved heterogeneity. The second dataset was first used by \citep{barreto1994effect}. It is characterized by the presence of multiple events because the individuals in the sample had at least one episode of diarrhea throughout their stay in the study. Our interest is to evaluate the effects of some covariates on the times between diarrhea recurrences. The use of frailties, in this case, aims to deal with a possible association between times until the recurring events. Applications of \texttt{readmission} and \texttt{diarrhea} will be referred to throughout the text as $\mathcal{A}_1$ and $\mathcal{A}_2$, respectively. For the inference procedure, in both applications, we choose weakly informative prior: $\pmb{\psi}, \pmb{\phi} \sim \mbox{Normal}~(0,3); \pmb{\gamma} \sim \mbox{LogNormal}~(0,2); \text{ and } \sigma_w \sim \mbox{Gamma}~(1,1)$. In all models whose baseline is PE or BP, we use $m = 5$. 

To compare the quality of the fits of the models in both applications, we applied the Widely Applicable Information Criterion (WAIC) criterion. The WAIC is a goodness-of-fit measure for statistical models, especially useful in Bayesian contexts \citep{ninomiya2021prior}. It is a generalization of the well-known Akaike Information Criterion (AIC) and is applicable even when the model is complex or when the number of parameters is large concerning the number of observations \citet{akaike2011akaike}. We applied the function \texttt{waic} available in the package \texttt{loo} \citep{vehtari2021package}. The WAIC calculation steps are in Appendix B. Among the frailty models fitted in applications $\mathcal{A}_1$ and $\mathcal{A}_2$ (PH$_{EX}$, PH$_{PE}$, PH$_{BP}$, PO$_{EX}$, PO$_{PE}$, PO$_{BP}$, YP$_{EX}$, YP$_{PE}$, and YP$_{BP}$), the primary interest is in the model with the best WAIC score, for which we will report estimates. The WAIC values obtained in applications $\mathcal{A}_1$ and $\mathcal{A}_2$ are available in Table \ref{tab:waic}.

\begin{table}[ht]
\centering
\caption{WAIC values obtained in applications $\mathcal{A}_1$ and $\mathcal{A}_2$}
\label{tab:waic}
\small
\begin{tabular}[t]{crr}
\toprule
          & \multicolumn{2}{c}{WAIC}                                                  \\
          \cline{2-3} 
Model     & \multicolumn{1}{c}{$\mathcal{A}_1$} & \multicolumn{1}{c}{$\mathcal{A}_2$} \\
\midrule
PH$_{EX}$ & 1832.3565                           &   2828.3798                                  \\
PH$_{PE}$ & 1834.1247                           & 2663.4610                                    \\
PH$_{BP}$ & 1819.1796                           &  2621.1369                                 \\
PO$_{EX}$ & 1837.6724                           &  2657.5942                                   \\
PO$_{PE}$ & 1838.0630                           & 2595.1940                                    \\
PO$_{BP}$ & 1829.8205                           & 2577.9906                                    \\
YP$_{EX}$ & 1835.2884                           & 2671.5956                                     \\
YP$_{PE}$ & 1837.6724                           & 2642.7094                                     \\
YP$_{BP}$ & \textbf{1818.9457}                  &  \textbf{2528.7280}                               \\
\bottomrule
\end{tabular}
\end{table}


Let's analyze the application $\mathcal{A}_1$. The dataset \texttt{readmission} originates from Bellvitge's Public University Hospital in Barcelona, Spain, capturing medical records from January 1996 to December 1998. The study focused on 403 individuals who underwent surgery. Some follow-up termination occurred in cases of patient death, migration, or hospital transfer. Only 103 patients died. Four time-fixed effects recorded in the file will be considered: (1) \texttt{sex} (\texttt{Male}, when \texttt{sex} = 0, or \texttt{Female}, when \texttt{sex}=1); (2) \texttt{chemo} which represents whether there was chemotherapy treatment (\texttt{Treated}, when \texttt{chemo}=1, or \texttt{nonTreated}, when \texttt{chemo}=0) and, (3) \texttt{dukes} which represents the Dukes' stage. \citet{gonzalez2005sex} classified the sample pacients as (\texttt{A-B}, \texttt{C} or \texttt{D}). 
Table \ref{tab:DummyDukes} shows how we configure two dummy variables to accommodate the three levels of Dukes' stages, \texttt{A-B}, \texttt{C}, and \texttt{D}.

\begin{table}[ht]
\centering
\caption{Dummy variable for variable \texttt{dukes}.}
\label{tab:DummyDukes}
\small
\begin{tabular}[t]{ccc}
\toprule
Dukes' stages       & Dukes$_1$	  & Dukes$_2$ \\          
\midrule
\texttt{A-B}		& 0		      & 0\\
\texttt{C}			& 1		      & 0\\
\texttt{D}			& 0		      & 1\\
\bottomrule
\end{tabular}
\end{table}

We observed that 164 were women, 239 were men; 217 received chemotherapy treatment and 186 did not. Colorectal cancer of 180 patients was classified as Dukes' stage \texttt{A-B}, of 148 as Dukes' stage \texttt{C}, and of 75 as Dukes' stage \texttt{D}.

In $\mathcal{A}_1$, we generated four MCMC chains for each parameter via \texttt{rstan} \citet{rstanMain} with 5000 iterations, of which 2500 are warm-ups, resulting in posterior samples of size 10000. This large volume of posterior samples aimed to allow a better convergence of $\sigma_w$. The model with the best WAIC is YP$_{BP}$. By this model, the estimate for the effect of the variable \texttt{sex} suggests that being female might be associated with a lower likelihood of death compared to being male, but this result is not statistically significant, since the credibility interval includes zero. 

The estimate for \texttt{sex} variable is negative, both in the short and long term. The negative value suggests that being female (\texttt{sex}=1) might be associated with a lower probability of death compared to being male (\texttt{sex}=0), but the credible interval includes 0, indicating this result is not statistically significant. The variable \texttt{chemo} is also not significant. 
The estimate for \texttt{Dukes}$_2$ is positive and signifcative, in the short terms. Its value indicates that being at Dukes' stage \texttt{D} in comparison to \texttt{A-B} is associated with a higher likelihood of death. 
The interaction between chemotherapy treatment and Dukes' stage is significant in the short term. This means that individuals with cancer in more advanced stages, even if they undergo chemotherapy at the beginning of follow-up, have a higher risk of death.
We note that the estimates of the parameter $\sigma_w$ suggest that there is variability among study individuals that is not explained by the model's fixed effects alone. The estimates of the other models can be seen by accessing \href{https://cassiushenrique.shinyapps.io/appRealFrailty/}{cassiushenrique.shinyapps.io/appRealFrailty}. 

\begin{table}[ht]
\centering
\caption{Summary of the YP$_{BP}$ model to the readmission data: posterior mean estimate (est), standard deviation (sd) along with the 95\% credibility interval (LW; UP).}
\label{tab:fitFra}
\small
\begin{tabular}{llrrrrr}
\toprule
\multicolumn{4}{c}{ } & \multicolumn{2}{c}{95\% CI}\\
\cmidrule(l){5-6}
\multicolumn{1}{c}{par} & \multicolumn{1}{c}{description} & \multicolumn{1}{c}{est} & \multicolumn{1}{c}{sd} & \multicolumn{1}{c}{LW} & \multicolumn{1}{c}{UP} \\ 
  \midrule
$\psi_1$ & Sex & -0.2122 & 0.4257 & -1.0333 & 0.6558 \\ 
$\psi_2$ & Chemo & -0.3307 & 0.7277 & -1.6513 & 1.2817 \\ 
$\psi_3$ & Dukes$_1$ & 1.1541 & 0.6554 & -0.1094 & 2.5243 \\ 
$\psi_4$ & Dukes$_2$ & \textbf{3.6413} & 0.5822 & 2.5000 & 4.8180 \\ 
$\psi_5$ & Chemo * Dukes$_1$ & \textbf{2.0182} & 0.9840 & 0.0650 & 3.9126 \\ 
$\psi_6$ & Chemo * Dukes$_2$ & \textbf{2.8088} & 0.9107 & 0.9140 & 4.5319 \\ 
\midrule
$\phi_1$ & Sex & -0.6734 & 1.1280 & -2.6184 & 2.0198 \\ 
$\phi_2$ & Chemo & -0.7680 & 2.0203 & -3.8750 & 3.6330 \\ 
$\phi_3$ & Dukes$_1$ & -1.1885 & 1.4296 & -2.7703 & 2.8973 \\ 
$\phi_4$ & Dukes$_2$ & 2.0736 & 1.6249 & -0.3909 & 5.8904 \\ 
$\phi_5$ & Chemo * Dukes$_1$ & 0.7192 & 2.1300 & -3.6591 & 4.4600 \\ 
$\phi_6$ & Chemo * Dukes$_2$ & 0.2049 & 2.1791 & -4.1632 & 4.1804 \\ 
\midrule
$\sigma_w$ & sd(Frailty) & {1.2043} & 0.4062 & 0.3946 & 1.9889 \\ 
  \bottomrule
\end{tabular}
\end{table}

\begin{figure}[ht]
    \centering
    \caption{Kaplan-Meier (step function) and survival curves estimated by YP$_{BP}$ model (continuous function) about the terminal event for the levels of variables (A) \texttt{sex}, (B) \texttt{chemo}, and (C) \texttt{dukes}.}
    \includegraphics[scale=0.3]{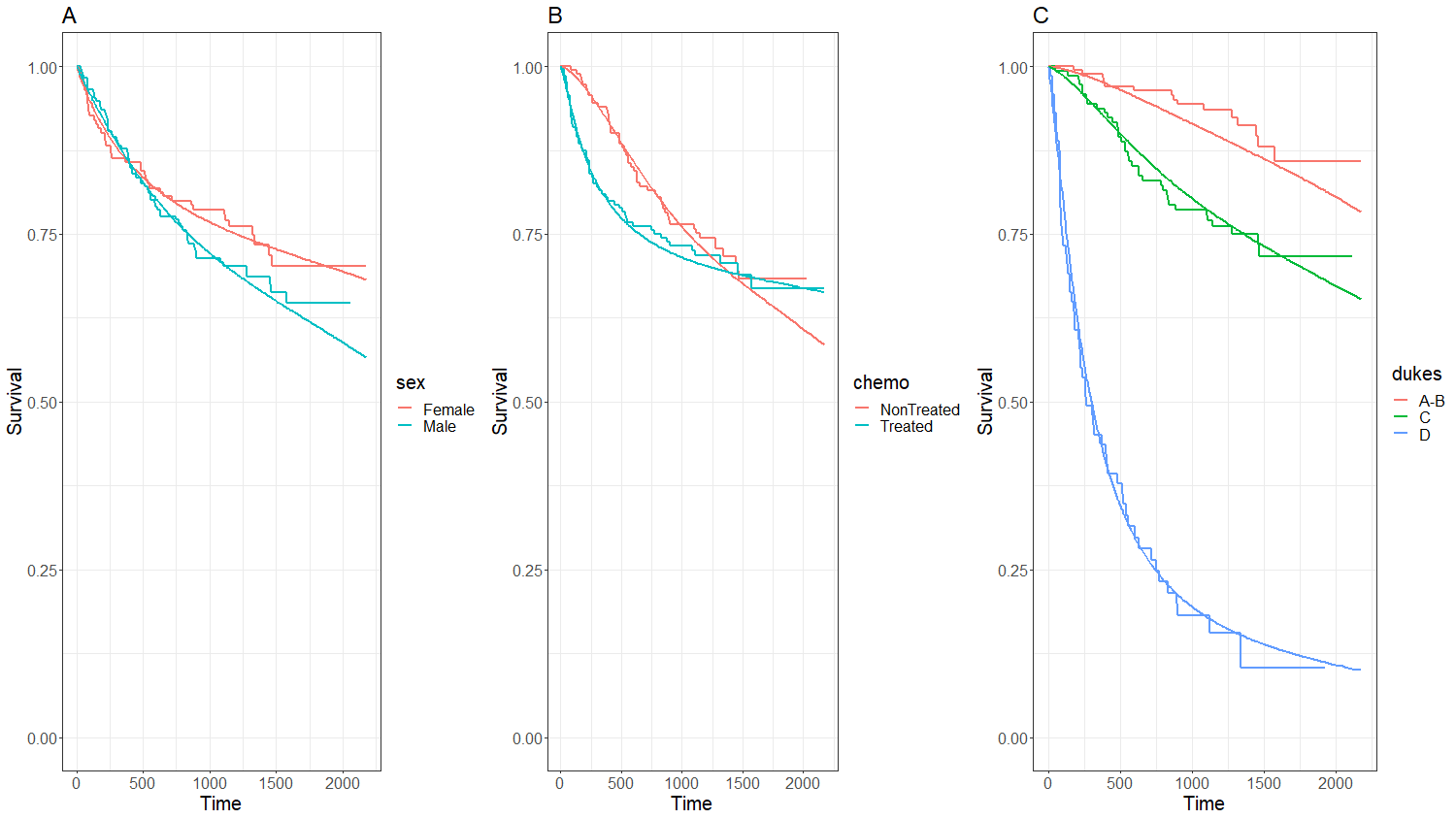}
    \label{fig:KMBPFra}
\end{figure}

From Figure \ref{fig:KMBPFra}, we can notice a similarity between the survival functions estimated by our model (represented by the continuous curves) and the Kaplan-Meier estimates. This fact suggests that our model is capturing the pattern of the observed data. However, is perceived the high volume of censoring in this dataset. Our models do not include cure fraction models. The inclusion of this approach could increase the precision of long-term effect estimates. Figure \ref{fig:KMBPFra}-A displays the survival curves for both men and women. This figure suggests that there is insufficient statistical evidence to conclude that a patient's sex significantly influences survival time.Figure \ref{fig:KMBPFra}-B presents the survival curve for patients treated or not treated with chemotherapy.
Our model estimates that the survival curves intersect for the first time on the day $1397$. Note that the survival curves cross more than once, however, the long-term effect tends to be zero. In Figure \ref{fig:KMBPFra}-C, we see the survival functions for Dukes' stages \texttt{A-B}, \texttt{C}, and \texttt{D} present well-defined differences between them. The Dukes' stage \texttt{D} shows the lowest survival functions over time, while patients with Dukes's stages \texttt{A-B} colorectal cancer appear to survive longer.

We now discuss the results of the application $\mathcal{A}_2$. The \texttt{diarrhea} dataset originates from a community-based, randomized, placebo-controlled study spearheaded by \citet{barreto1994effect}, conducted from December 1990 to December 1991, and supported by the Institute of Collective Health of the Federal University of Bahia. This research aimed to evaluate the effects of vitamin A supplementation on diarrhea incidence among children aged 6 to 48 months in the Northeast of Brazil. The participating children had an average age of 29.32 months, with a standard deviation of 12.18 months and a median age of 30 months. 

This dataset comprises 860 children, categorized into two groups based on the covariate \texttt{treatment}: 426 in the placebo group (\texttt{treatment} = 0) and 434 who received vitamin A supplementation (\texttt{treatment} = 1). In the sample, there were 403 females (\texttt{sex} = 0) and 457 males (\texttt{sex} = 1). Given the occurrence of one or more diarrhea episodes among various participants during the study period, the dataset encompasses 5592 records. On average, the children experienced 6.502 episodes of diarrhea, with the median frequency at 5 episodes. The number of recurrences varied from a minimum of 1 to a maximum of 27 with a standard deviation of 5.24.

The dataset recorded five covariates, three of which are constant (\texttt{treatment}, \texttt{sex}, and \texttt{age}) and were documented at the onset of the study. The remaining two covariates, which vary per episode, include the number of days without diarrhea before the current episode and the average number of liquid or semi-liquid stools from the preceding diarrhea episode. However, these time-dependent covariates are not considered in this particular analysis due to the limitations of the models discussed, which cannot accommodate such variables.

In $\mathcal{A}_2$, four MCMC chains for each parameter via \texttt{rstan} \citet{rstanMain} are also generated, each with 2000 iterations, of which 1000 are warm-ups, resulting in subsequent samples of size 1000. We report the estimates from YP$_{BP}$ here because this model has the best WAIC score. The fit results are in Table \ref{tab:fitFraDiarreia}. The estimates of the other models can be seen online \href{https://cassiushenrique.shinyapps.io/appRealFrailtyDiarrhea/}{https://cassiushenrique.shinyapps.io/appRealFrailtyDiarrhea}. The treatment is only significant in the long term because the credibility interval for $\phi_1$ does not include zero. Over the long term, the treatment tends to reduce the risk of new diarrhea episodes. The sex of the individual is not significant either in the short or long-term concerning diarrhea occurrences, because the credibility intervals do not include zero. The age of the individual seems to be significant both in the short and long term, in that older children tend to have a lower risk of diarrhea recurrence both at the start and end of the follow-up. The standard deviation of frailty is significant, indicating that there is some association between the recurrent events.

\begin{table}[ht]
\centering
\caption{Summary of the YP$_{BP}$ model to the diarrhea data: posterior mean estimate (est), standard deviation (sd) along with the 95\% credibility interval (LW; UP).}
\label{tab:fitFraDiarreia}
\small
\begin{tabular}{llrrrrr}
\toprule
\multicolumn{4}{c}{ } & \multicolumn{2}{c}{95\% CI}\\
\cmidrule(l){5-6}
\multicolumn{1}{c}{par} & \multicolumn{1}{c}{description} & \multicolumn{1}{c}{est} & \multicolumn{1}{c}{sd} & \multicolumn{1}{c}{LW} & \multicolumn{1}{c}{UP} \\ 
  \midrule
   $\psi_1$ & treatment & \textbf{-0.1741} & 0.0839 & -0.3375 & -0.0070 \\ 
   $\psi_2$ & sex & 0.1323 & 0.0824 & -0.0262 & 0.2958 \\ 
   $\psi_3$ & Age(beginning) & \textbf{-0.0177} & 0.0030 & -0.0236 & -0.0120 \\ 
   \midrule
   $\phi_1$ & treatment & -0.1290 & 0.0646 & -0.2537 & 0.0016 \\ 
   $\phi_2$ & sex & -0.0019 & 0.0640 & -0.1211 & 0.1272 \\ 
   $\phi_3$ & Age(beginning) & \textbf{-0.0368} & 0.0023 & -0.0412 & -0.0324 \\ 
   \midrule
   $\sigma_w$ & sd(Frailty) & \textbf{0.6350} & 0.0293 & 0.5793 & 0.6934 \\
   \hline
\end{tabular}
\end{table}

%% file: chapters/6-Conclusion.tex
This work proposed to develop a class of models within a Bayesian framework, designed to explain the impact of observed characteristics on survival curves that may intersect. For this finally, we used the YP regression structure for its ability to encompass and generalize the PH and PO models. The class of models embraces YP frailty. The incorporation of frailty in these models constitutes a contribution of this study, since in the literature the YP models did not incorporate frailty. We combined exponential, PE, and BP baseline functions. The selection of these last two baseline functions was motivated by their versatility because can fit a variety of hazard function shapes. In that regard, the innovations promoted by this work are the YP$_{EX}$, YP$_{PE}$, YP$_{BP}$ frailty models.

The models of the class enable the analysis of survival data under distinct scenarios: ($\mathcal{S}_1$) individuals with a unique survival time where individual frailty explains unobserved heterogeneities; ($\mathcal{S}_2$) individuals who experience recurring events for which the shared frailty accommodates the association between the survival times of the same individual. 

As for numerical results, we executed a Monte Carlo simulation study for the class aimed at evaluating the influence of model selection on parameter estimation. This assessment focused on some criteria such as estimation biases (RB), average standard error (ASE), standard deviation of estimates (SDE), credibility intervals, and coverage probability (CP). A total of $M_C = 250$ Monte Carlo replicas were generated, each comprising $L = 300$ individuals. In the two scenarios, our estimates mean and median are generally close to the true values, indicating a good level of accuracy. In addition, the ASE and SDE values are close and the CP values are not very far from 95\%. These facts signal a good performance of our models. In the real application, we fitted our models on the \texttt{readmission} and \texttt{diarrhea} databases and we report the estimates and interpretations of our models with the best WAIC.

This research has some limitations. In our simulation studies, we did not apply WAIC to Monte Carlo samples. We acknowledge that the assessment of these values could enhance the depth of comparative analysis of our models. We did not consider Weibull baseline distributions in the fit of our models. In the real application of \texttt{readmission}, we believe that the high volume of administrative censoring somewhat reduced the predictive ability of our models. Our models are not yet capable of accommodating time-dependent variables. In future research, we want to apply the following approaches: (A) Evaluate the WAIC of our model fits in a simulation study. (B) Conduct a more extensive simulation study incorporating the baseline Weibull distribution. (C) Incorporate a cure fraction model into our model classes. (D) Adapt our regression frameworks to allow us to model time-dependent covariates. (E) Extend our models to a frequentist approach. (F) Deploy a residual analysis that provides an additional way of evaluating the quality of our fits. (G) Publish an R package that provides the functions used in this work, facilitating the replication of its results.